\documentclass{appolb}
\usepackage{epsfig}

\begin{document}
\title  {{The optical model potential \\ of the $\Sigma$
hyperon in nuclear matter}\footnote{Presented at the XXXI Mazurian
Lakes Conference on Physics, Piaski, Poland, August 30--September
5, 2009}
\author { Janusz Dabrowski and Jacek Rozynek}
\address {Theoretical Physics Division, So{\l}tan Institute
for Nuclear Studies,\\
Ho{\.z}a 69, 00-681 Warsaw, Poland}}

\maketitle

\begin{abstract}

We present our attempts to determine the optical model potential
$U_\Sigma = V_\Sigma -iW_\Sigma$ of the $\Sigma$ hyperon in
nuclear matter. We analyze the following sources of information
on $U_\Sigma$: $\Sigma N$ scattering, $\Sigma^-$ atoms, and final state
interaction of $\Sigma$ hyperons in the $(\pi,K^+)$ and $(K^-.\pi)$
reactions on nuclear targets. We conclude that $V_\Sigma$ is repulsive
inside the nucleus and has a shallow a tractive pocket at the nuclear
surface. These features of $V_\Sigma$ are consistent with the
Nijmegen model F of the hyperon-nucleon interaction.

\end{abstract}


\section{Introduction}

The interaction of the $\Sigma$ hyperon with nuclear matter may be
represented by the complex single particle (s.p.) optical model
potential $U_\Sigma=V_\Sigma-iW_\Sigma$. In this paper we present
our attempts to determine $V_\Sigma$ and $W_\Sigma$. We also point
out the most realistic two-body $\Sigma N$ interaction among the
available OBE models of the baryon-baryon interaction.

In the present paper we discuss the
following sources of information on $U_\Sigma$:
$\Sigma N$ scattering data in Sec.2, $\Sigma^-$ atoms
in Sec.3, associated production reactions in Sec.4,
and strangeness exchange reactions in Sec.5. Our conclusions
are presented in Sec.6.

\section{$\Sigma N$ scattering}

The way from the $\Sigma N$ scattering data to $U_\Sigma$ consists
of two steps: first, we determine the two-body $\Sigma N$ interaction
${\cal V}_{\Sigma N}$, and second, with this ${\cal V}_{\Sigma N}$
we calculate $U_\Sigma$. The scarcity of the two-body $\Sigma N$
data makes the first step
very difficult. A way of overcoming these difficulties was followed
by de Swart and his collaborators in Nijmegen: they assumed the
mechanism of one-boson exchange (OBE) and the SU(3) symmetry
which enabled them to employ the numerous $NN$ data in determining
the parameters of their two-body interaction. In this way they
produced a number of the Nijmegen models of the baryon-baryon
interaction: models D \cite{D}, F \cite{F}, soft core (SC) model
\cite{SC}, and the new soft-core (NSC) model \cite{NSC}.
\begin{figure}[h]
\vspace{1mm}
\begin{minipage}[t]{0.475\linewidth}
\centering \vspace{-8mm} {\psfig{file=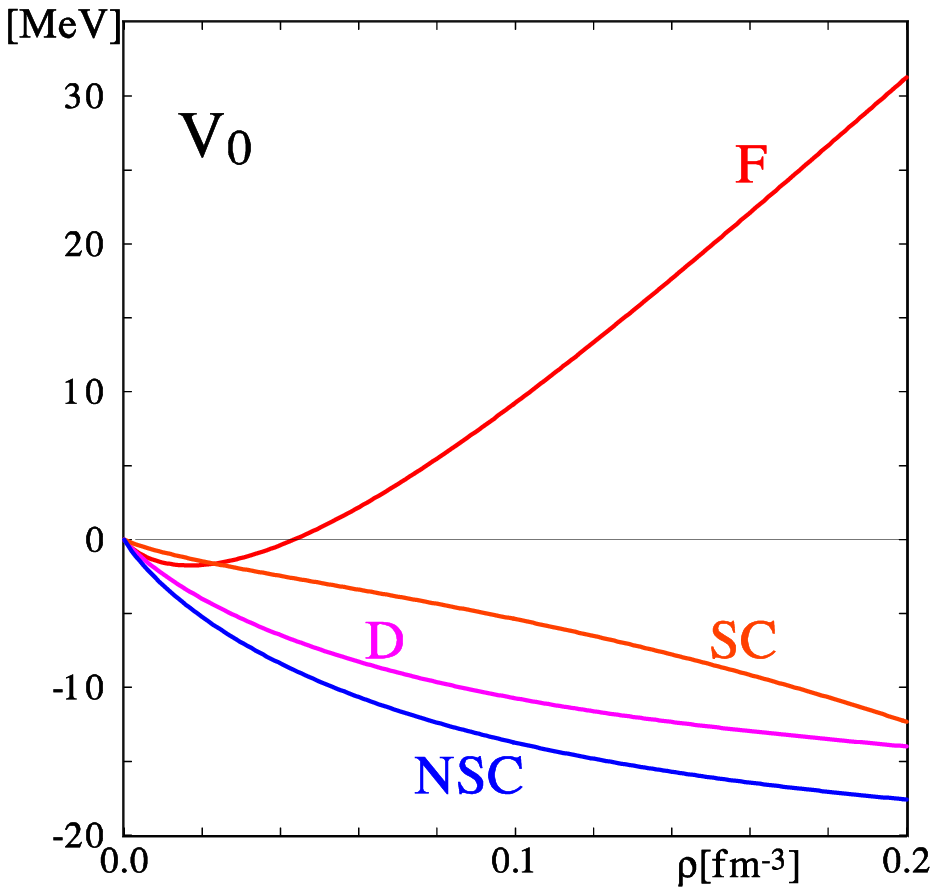,width=7cm}}
\caption{The isoscalar potential $V_\Sigma$ as a function of the
nucleon density $\rho$ at $k_\Sigma=0$ for the indicated Nijmegen
models of the $\Sigma N$ interaction.}
\end{minipage}
\vspace{20mm} \hspace{1mm}
\begin{minipage}[t]{0.475\linewidth}
\centering \vspace{-8mm}
\includegraphics[height=5.5cm,width=8.cm]{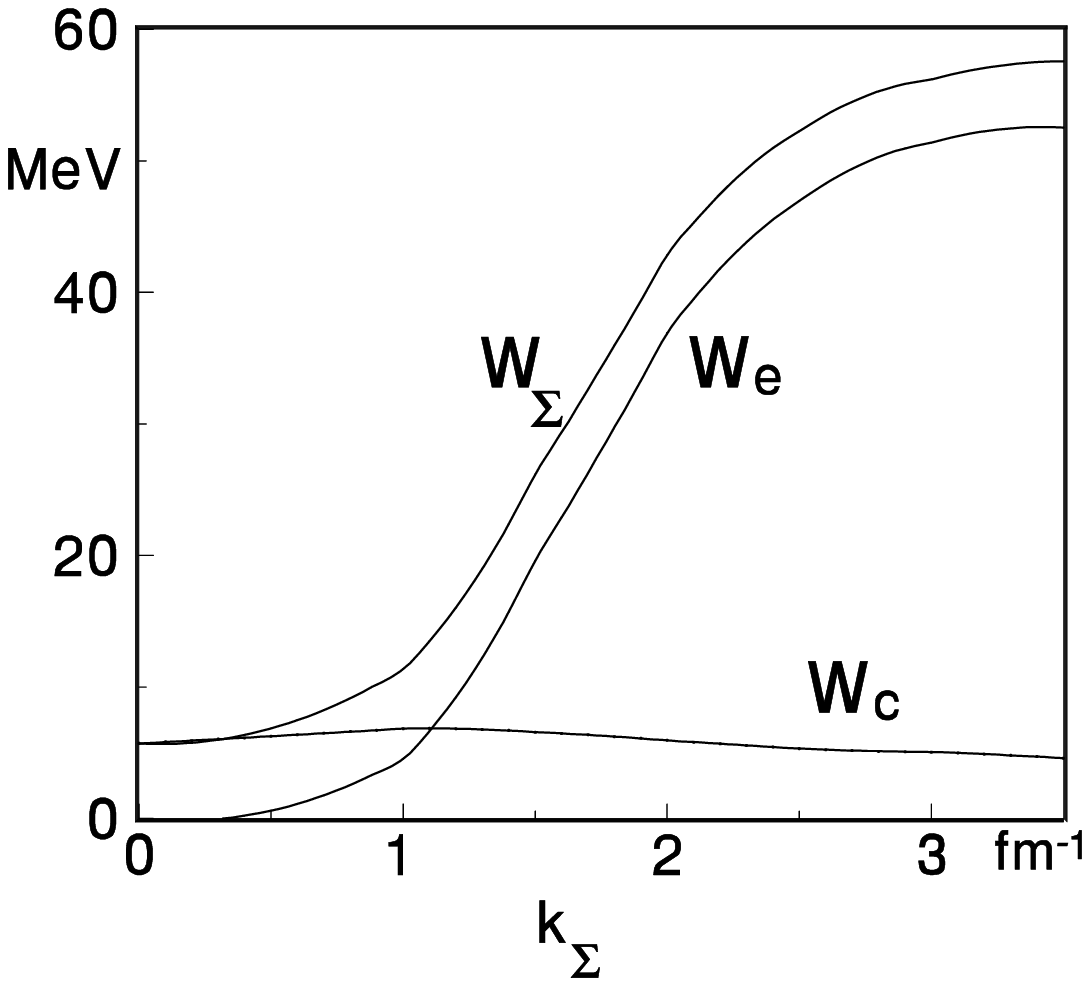}
\vspace{-6mm} \caption{The component $W_c,  W_\e$, and $W_t$ of
the $\Sigma$ absorptive potential in nuclear matter of density
$\rho_0$ as functions of $k_\Sigma$.}
\end{minipage}
\vspace{-16mm}
\end{figure}
\vspace{-8mm}
\subsection{The real potential $V_\Sigma$}

In calculating $V_\Sigma$ we use the real part of the effective
$\Sigma N$ interaction YNG \cite{YM} in nuclear matter. The YNG interaction
is the configuration space representation of the G matrix calculated
in the low order Brueckner approximation with the Nijmegen
models of the baryon-baryon interaction. Our results obtained for
$V_\Sigma$ as function of the nucleon density
$\rho$ are shown in Fig. 1. As the dependence of
$V_\Sigma$ on the $\Sigma$ momentum $k_\Sigma$ is not very strong
in the relevant interval of $k_\Sigma$ \cite{JJDD}, we use for
$V_\Sigma$ its value calculated at $k_\Sigma=0$.
We see that all the Nijmegen interaction models, except for model
F, lead to pure attractive $V_\Sigma$ which implies the existence of
bound states of $\Sigma$ hyperons in the nuclear core, \ie,
$\Sigma$ hypernuclei. Since no $\Sigma$ hypernuclei have been
observed,\footnote{The observed bound state of $^4_\Sigma$He
\cite{He} is an exception. In the theoretical description
of this state, Harada and his collaborators \cite{Ha} apply
phenomenological $\Sigma N$ interactions, in particular, the
interaction SAP-F simulating at low energies the Nijmegen
model F interaction. They show that essential for the existence
of the bound state of $^4_\Sigma$He is a strong Lane component
$V_\tau$ in $V_\Sigma$, and among the Nijmegen models the
strongest $V_\tau$ is implied by model F.\cite{JDJD}}
we conclude that among the Nijmegen interaction models
model F is the only realistic representation of the $\Sigma N$
interaction.

\subsection{The absorptive potential $W_\Sigma$}

As pointed out in \cite{YM}, the imaginary part of the YNG
interaction is very sensitive to the choice of the intermediate
state energies in the $G$ matrix equation. In this situation
we decided to use for $W_\Sigma$ the semi-classical
expression in terms of the total cross sections (modified by
the exclusion principle) for $\Sigma N$ scattering,
described in \cite{JDPR}. We denote by $W_c$ the contribution to the
absorptive potential of
the $\Sigma \Lambda$ conversion process $\Sigma N
\rightarrow \Lambda N^\prime$ and by $W_e$ the contribution
of the $\Sigma N$ elastic scattering, and have
$W_\Sigma = W_t = W_c + W_e$.
\footnote{
Notice that in the case of the nucleon optical potential
in nuclear matter (for nucleon energies below the threshold
for pion production),
$V_N-iW_N$, only the elastic $NN$ scattering contributes to
$W_N$, and the situation is similar as in the case of the
contribution $W_e$ to $W_\Sigma$.}


Our results obtained for $W_c, W_e, W_t$ for nuclear matter (with
N=Z) at equilibrium density $\rho = \rho_0 =$ 0.166 fm$^{-3}$ are shown in
Fig. 2.
With increasing momentum $k_\Sigma$ the $\Sigma\Lambda$ conversion
cross section decreases, on the other hand the suppression of $W_c$ by the
exclusion principle weakens. As the net result
$W_c$ does not change very much with $k_\Sigma$. The same two mechanisms
act in the case of $W_e$. Here, however, the action of the exclusion
principle is much more pronounced: at $k_\Sigma=0$ the suppression of
$W_e$ is complete. At higher momenta, where the Pauli blocking is not
important, the total elastic cross section is much bigger than the
conversion cross section, and we have $W_e>>W_c$, and consequently
$W_\Sigma>>W_c$.

\section{$\Sigma^-$ atoms}

The available data on strong interaction effects in $\Sigma^-$ atoms
consist of 23 data points: strong interaction shifts $\epsilon$ and
widths $\Gamma$ of the observed levels. These shifts and widths can
be measured directly only in the lowest $\Sigma^-$ atomic levels.
The widths of the next to the last level can be obtained indirectly
from measurements of the relative yields of X-rays.

In \cite{JRA}, we have estimated the 23 values of $\epsilon$ and
$\Gamma$ from the difference between the eigenvalues of the
Schr\"{o}dinger equation of $\Sigma^-$ in $\Sigma^-$ atoms with
the strong $\Sigma^-$-atomic nucleus interaction and without
this interaction. To obtain this strong interaction, we
applied the local density approximation, and used our
optical model of Sec. 2. The agreement of our results,
calculated with the optical potentials (obtained with the
4 Nijmegen $\Sigma N$ interaction models)
with the 23 empirical data points is characterized by the
following values of $\chi^2$: $\chi^2$(model D) $>$ 130,
$\chi^2$(model F) = 38.1, $\chi^2$(model SC) = 55.0,
 $\chi^2$(model NSC) $>$ 904, and we conclude that the $\Sigma^-$
atomic data point out at model F as the best representation
of the $\Sigma N$ interaction.\footnote{Notice that the positive
sign of the measured values of $\epsilon$ requires an attractive
$\Sigma$ potential at the nuclear surface, \ie at low densities.}

\section{The associated production reactions}

The first associated $\Sigma$ production reaction $(\pi^-,K^+)$
was observed at KEK on $^{28}$Si target at pion momentum of
1.2 GeV/c  (\cite{anna1},\cite{anna3}), and this reaction is
the subject of the present analysis. We consider the reaction
$(\pi^-,K^+)$ in which the pion $\pi^-$ with momentum
${\bf k}_\pi$ hits a proton in the $^{28}$Si target in the state $\psi_P$
and emerges in the final state as kaon $K^+$ moving in the direction
$\hat{k}_K$ with energy $E_K$, whereas the hit proton emerges in the final
state as a $\Sigma^-$ hyperon with momentum $\bf{k}_\Sigma$.
We apply the simple
impulse approximation described in \cite{I2}, with $K^+$ and $\pi^-$ plane
waves, and obtain:
\begin{equation}
\label{ia}
d^3\sigma/d\hat{k}_\Sigma d\hat{k}_K dE_K\sim|\int d{\bf r}\exp(-i{\bf qr})
\psi_{\Sigma,{\bf k}_\Sigma}({\bf r})^{(-)*}\psi_P({\bf r})|^2,
\end{equation}
where the momentum transfer ${\bf q}={\bf k}_K-{\bf k}_\pi$, and
$\psi_{\Sigma,{\bf k}_\Sigma}({\bf r})^{(-)}$ is the $\Sigma$ scattering
wave function which is the solution of the s,p. Schr\"{o}dinger equation
with the s.p. potential
\begin{equation}
\label{us}
U_\Sigma(r)=(V_\Sigma-iW_\Sigma)\theta(R-r),
\end{equation}
where for $V_\Sigma$ and $W_\Sigma$ we use the nuclear matter results
discussed in Section 2, calculated at $\rho=n/[(4\pi/3)R^3]$, where
$n$=27 is the number of nucleons in the final state.

For the $^{28}$Si target nucleus we assume a simple shell model
with a square well s.p, potential $V_P(r)$ (which determines $\psi_P$)
with the radius $R_P$ (and with a spin-orbit term).
The parameters of $V_P(r)$ are adjusted to the proton separation
energies (in particular $R_P=3.756$fm). For $R$ we make the simple
and plausible assumption: $R=R_P$.

In the inclusive KEK experiments \cite{anna1}-\cite{anna3} only the energy
spectrum of kaons at fixed $\hat{k}_\Sigma$ was measured.
To obtain this
energy spectrum, we have to integrate the cross section (\ref{ia}) over
$\hat{k}_\Sigma$.

\begin{figure}[h]
\vspace{0mm}
\begin{minipage}[t]{0.475\linewidth}
\centering \vspace{-8mm} {\psfig{file=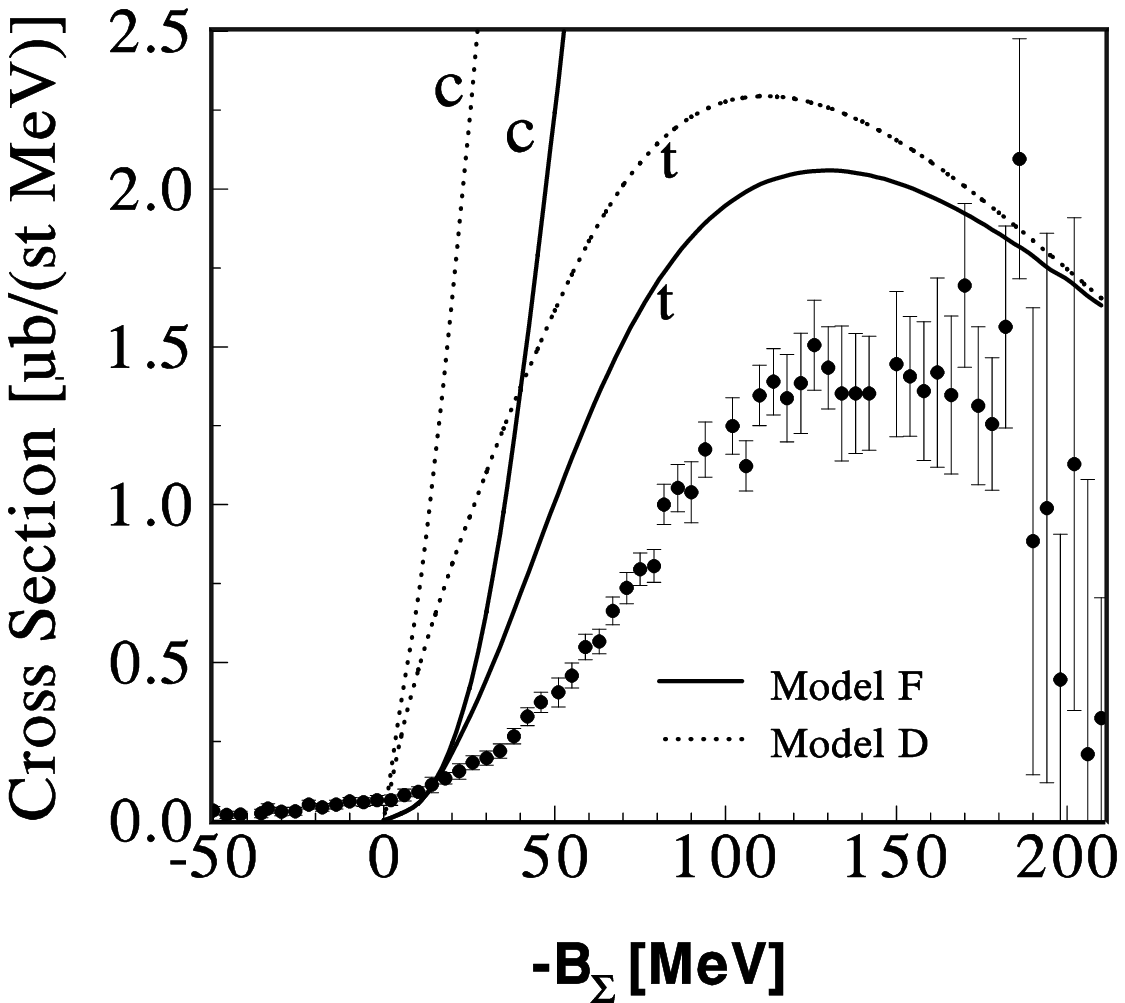,width=7.cm}}
\caption{Kaon spectrum from $(\pi^-,K^+)$ reaction on $^{28}$Si at
$\theta_K=6^\circ$ at $p_\pi=1.2$ GeV/c obtained with $V_\Sigma$
determined by models F and D of the $\Sigma N$ interaction. Curves
denoted by $c(t)$ were obtained with $W_\Sigma = W_c(W_t)$. Data
points are taken from \cite{anna3}.}
\end{minipage}
\vspace{20mm} \hspace{1mm}
\begin{minipage}[t]{0.475\linewidth}
\centering \vspace{-1mm}
\includegraphics[height=4.9cm,width=5.6cm]{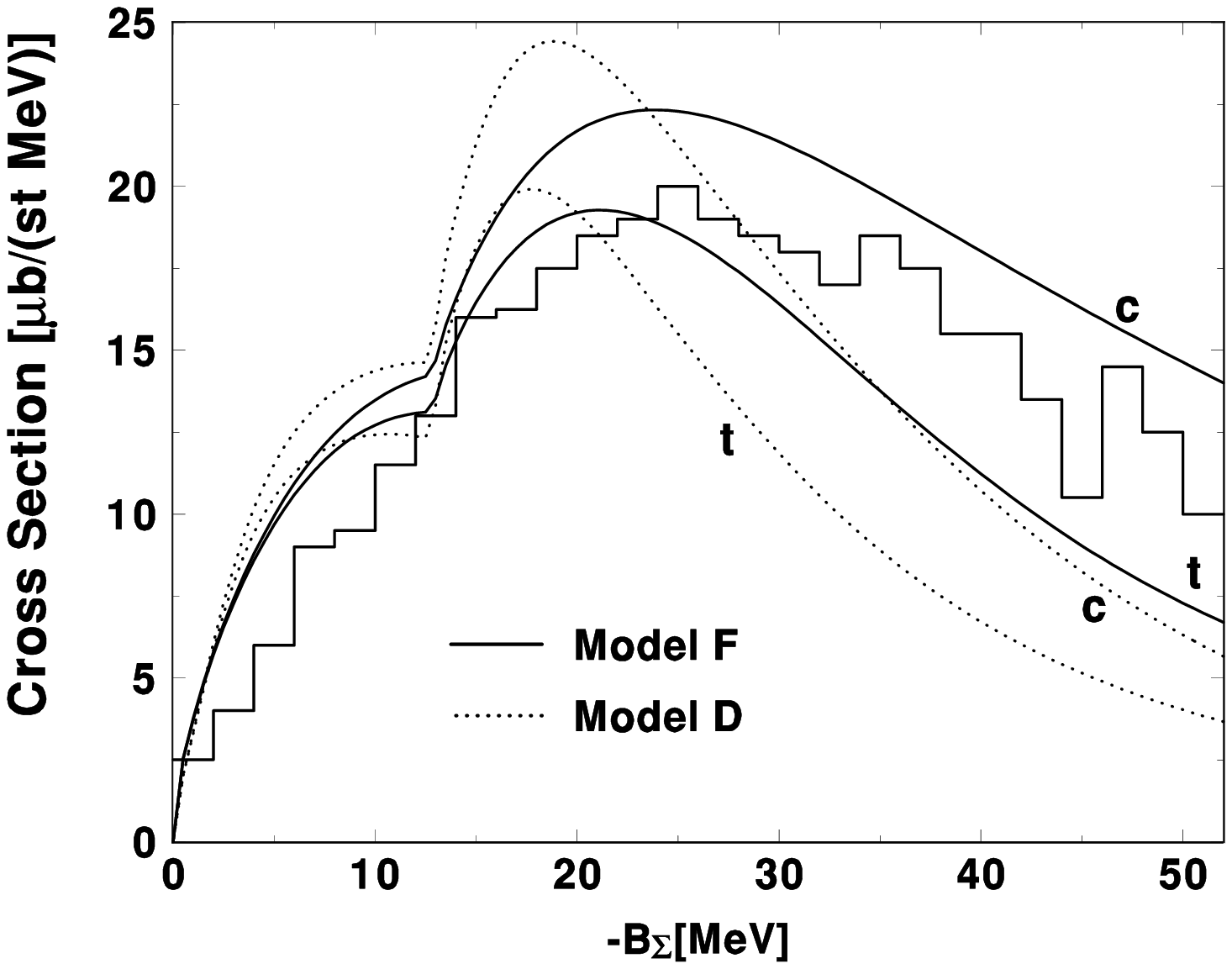}
\vspace{-3mm} \caption{Pion spectrum from $(K^-,\pi^+)$ reaction
on $^9$Be at $\theta_\pi=4^\circ$ at $p_K=0.6$ GeV/c obtained with
$V_\Sigma$ determined by models F and D of the $\Sigma N$
interaction. Curves denoted by $c(t)$ were obtained with $W_\Sigma
= W_c(W_t)$. Data points are taken from \cite{bart}.}
\end{minipage}
\vspace{-16mm}
\end{figure}
We present our results for the inclusive cross
section as a function of $B_\Sigma$, the separation (binding)
energy of $\Sigma$ from the hypernuclear system produced.
Our model F and D results
\footnote{The remaining models SC and NSC are
similar to model D: they all lead to attractive $V_\Sigma$ in
contradistinction to model F leading to repulsive $V_\Sigma$ (at
densities inside nulei - see Fig. 1). Consequently, the results
for the kaon spectrum for models SC and NSC are expected to be
similar as in case of model D.}
for kaon spectrum from $(\pi^-,K^+)$ reaction on $^{28}$Si
at $\theta_K=6^o$ at $p_\pi= 1.2$ GeV/c
are shown in Fig. 3.
We see that the best fit to the data points is obtained for
$V_\Sigma$ derived from model F and with $W_\Sigma=W_t=W_c+W_e$.
The fit would improve if we considered the distortion of kaon and
especially of pion waves
(it was noticed already in Ref. \cite{anna1} that this
distortion pushes the kaon spectrum down). Inclusion into the
absorptive potential of the contribution $W_e$ of the elastic
$\Sigma N$ scattering is essential for obtaining this result with
$V_\Sigma$(model F) = 17.25 MeV. Earlier estimates of the kaon
spectrum without this contribution suggested a repulsive
$V_\Sigma$ with an unexpected strength of about 100 MeV. Notice
that the action of the absorptive potential $W_\Sigma$ on the
$\Sigma$ wave function (decrease of this wave function) is similar
as the action of a repulsive $V_\Sigma$. Therefore we achieve with
strong absorption the same final effect with a relatively weaker
repulsion.

\section{The strangeness exchange reactions}

First observations of the strangeness exchange $(K^-,\pi)$ reactions
with a reliable accuracy were performed at BNL. Here,
we shall discuss the $(K-,\pi^+)$ reaction observed at BNL on
Be$^9$ target with 600 MeV/c kaons.\cite{bart} Proceeding similarly
as in the case of the associated production described in Sec.4, we
get the results shown in Fig. 4. We see that similarly as in Sect. 4
the fit to the data points obtained for $V_\Sigma$
derived from model F is much better than the fit obtained with
model D.

\newpage

\section{Conclusions}

$\bullet$ The real part $V_\Sigma$ of the $\Sigma$ optical potential
is repulsive inside the nucleus and has a shallow attractive pocket
at the nuclear surface.

$\bullet$ Among the Nijmegen models of the baryon-baryon interaction
only model F leads to this form of $V_\Sigma$.

$\bullet$ The contribution of the elastic $\Sigma N$ scattering to the
absorptive part $W_\Sigma$ of the $\Sigma$ optical potential is
essential in the analysis of $\Sigma$ production processes.
\vspace{0.6cm}

This research was partly supported by the Polish Ministry of Science
and Higher Education under Research Project No. N N202 046237.

\newpage

\end{document}